\documentstyle[aaspp4]{article}
 
\newcommand{\Lsun}{\mbox{$L_{\sun}$}}

\newcommand{\etal}{et al.~}
\newcommand{\um}{\mbox{$\,\mu\rm m$}}

\begin{document}

\title{
The Mid-Infrared Spectra of Normal Galaxies\\
}

\vspace{0.20in}
\author{G. Helou and Nanyao Y. Lu}
\affil{Infrared Processing and Analysis Center,\\
        California Institute of Technology, \\
        MS 100-22, Caltech\\
        Pasadena, CA 91125\\
        Emails: gxh@ipac.caltech.edu; lu@ipac.caltech.edu}

\author{M. W. Werner}
\affil{ Jet Propulsion Laboratory, \\
        California Institute of Technology, \\
        4800 Oak Grove Drive\\
	Pasadena, CA 91109\\
	Email: mww@ipac.caltech.edu}

\author{S. Malhotra and N. Silbermann}
\affil{Infrared Processing and Analysis Center\\
        MS 100-22, Caltech\\
        Pasadena, CA 91125\\
	Emails: san@ipac.caltech.edu; nancys@ipac.caltech.edu}
\vspace{0.25in}

\begin{abstract}
\indent The mid-infrared spectra (2.5 to 5 and 5.7 to 11.6\um) obtained by
ISO-PHOT reveal the interstellar medium emission from galaxies powered by
star formation to be strongly dominated by the aromatic features at 6.2,
7.7, 8.6 and 11.3\um.
Additional emission appears in-between the features, and an underlying
continuum is clearly evident at 3--5\um.  This continuum would contribute 
about a third of
the luminosity in the 3 to 13\um \ range.  The features together carry
5 to 30\% of the  40-to-120\um ``FIR'' luminosity.  The relative
fluxes in individual features depend very weakly on galaxy parameters such
as the far-infrared colors, direct evidence that the emitting particles are
not in thermal equilibrium.  The dip at 10\um\, is unlikely to result from
silicate absorption, since its shape is invariant among galaxies.  
The continuum component has a $\rm
f_\nu\propto \nu^{+0.65}$ shape between 3 and 5\um, and carries 1 to 4\% of the
``FIR'' luminosity; its  extrapolation to longer wavelengths falls 
well below  the spectrum in the 6 to 12\um\, range. This continuum component
is almost certainly of non-stellar origin, and is probably due to
fluctuating grains without aromatic features.  The spectra reported here
typify the integrated emission from the interstellar medium of the majority
of star-forming galaxies, and could thus be used to obtain redshifts 
of highly extincted galaxies up to z=3  with SIRTF.

\end{abstract}


\section{Introduction} \label{sec1}

The Infrared Space Observatory (ISO; Kessler
et al. 1996) has provided a unique opportunity for 
infrared spectroscopy at wavelengths
and sensitivities inaccessible to sub-orbital platforms.  Mid-infrared
spectroscopy has been an important tool in characterizing star formation 
and the interstellar medium in galaxies since
the mid-seventies (Willner \etal 1977; Roche \etal 1991),
and has taken a major leap forward thanks to the
sensitivity and unimpeded spectral coverage of ISO.

This is a first report on mid-infrared spectroscopy of galaxies using
ISO-PHOT (Lemke \etal 1996) obtained as part of the ISO Key Project under
NASA Guaranteed Time on the interstellar medium of normal galaxies (Helou
\etal 1996).  This Key Project used ISO to derive the physical properties
of the interstellar gas, dust and radiation field in a broad sample of
``normal'' galaxies, defined as systems whose luminosity is derived
from stars.  This sample includes
sixty objects comprising all morphological types, with visible-light
luminosities ranging from $10^8$ to $10^{11}$ \Lsun, infrared-to-blue
ratios from 0.1 to 100, and IRAS colors
$R(60/100)=f_{\nu}(60)/f_{\nu}(100)$ between 0.3 and 1.2.  The sample is
not statistically complete, but is designed to capture the
great diversity among galaxies, especially in terms of the ratio of current
to long-term average star formation rate.

\section{The Spectra} \label{sec2}

The PHT-S module of the ISO-PHOT instrument  (Lemke \etal 1996) has a
$24^{\prime\prime}\times24^{\prime\prime}$ aperture on the sky, pointed
with an accuracy $\leq2^{\prime\prime}$ (Kessler \etal 1996).  The
instrument has two 64-element linear Si:Ga detector arrays covering the
range 2.5 -- 4.9\um\ with $\Delta\lambda=0.04\um$ per element,
and the range 5.9 -- 11.7\um\ with 0.1\um\ per element.
The elements are sized to match the image of the entrance aperture, 
thereby  determining the spectral resolution.  The FWHM
of an unresolved line varies between 1.5 and
2 elements depending on the centering of the line with respect to pixel
boundary. 
Each galaxy was observed for a total of 512 seconds, split evenly 
between galaxy and sky, using a double-sided chopping scheme for sky 
subtraction.  
The PHT-S spectra were derived from the Edited Raw Data using the
ISO-PHOT Interactive Analysis (PIA) V.7 in a standard way. 
The flux calibration was done using a mean, signal-dependent ``detector
response function'' derived directly from chopped observations of standard
stars with known spectra.  Our final spectra are 
the integral under the PHT-S beam profile of the
surface brightness distribution of the source, expressed as a
flux density.  The combined uncertainties
of the relative calibration across the spectrum and the absolute flux scale
should be $\leq$30\% according to the PHT-S calibration report as well as
our own cross-calibration with the CAM photometry at 6.7\um\, in Silbermann
\etal (2000).

Of the 45 galaxies eventually observed in total with PHT-S, figure 1 shows
spectra for seven galaxies selected so that most of their flux is contained
within the PHT-S aperture (Table 1).  This selection is based on broadband
images at 6.75\um\ ($\Delta\lambda\simeq3\um$) obtained with ISO-CAM
(C\'esarsky et al 1996) and described elsewhere (Silbermann et al 2000).
Table 1 lists the galaxies and illustrates the large spread in their 
basic properties.
Column (1) gives the name, (2) the IRAS color ratio $R(60/100)$,
and (3) the optical morphology.  Column (4) gives the fraction of 6.75$\um$ \
flux within the PHT-S aperture, (5) the infrared-to-blue ratio, (6) the
luminosity in solar units within the FIR synthetic band (42.5 to 122.5\um)
defined in Helou \etal (1988), and (7) the mid-infrared morphology
from Silbermann \etal (2000).

The mid-infrared spectra of all these galaxies are dominated 
by emission features which appear in two main groups,
one stretching from 5.5 to 9\um, and the other starting at 11\um\ and
extending beyond the spectral range of these data (see Boulanger \etal 1996
for a similar spectrum at longer wavelengths of a molecular cloud region).
The shape and relative strengths of the features are quite similar to
``Type A sources'' which are the most common non-stellar objects in the
Milky Way: reflection nebulae, planetary nebulae, molecular clouds, diffuse
atomic clouds, and HII regions (Geballe 1997, Tokunaga 1997, and references
therein).  While there is good evidence to link these features to
Polycyclic Aromatic Hydrocarbons (PAH) or similar compounds, there is no
rigorous spectral identification (Puget \& L\'eger 1989; Allamandola
\etal 1989).  It is generally agreed however that the emitters are  
small structures, $\sim$100 atoms typically,
transiently excited to high internal energy levels by single photons.
The identification issue will not be discussed further here, and the spectral
features will be referred to as ``aromatic features in emission"  (AFE).   
Quantitatively
similar spectra have been reported from spectroscopic observations with 
PHT-S, ISO-CAM CVF or ISO-SWS 
on a variety of Galactic sources 
and a number of galaxies (Tielens \etal 1999 review, Vigroux \etal 1996,
Metcalfe \etal 1996).  
However, interstellar dust can manifest mid-infrared spectra of 
fundamentally different appearance 
in environments such as the Galactic Center (Lutz \etal 1996), supernova
remnants (Tuffs 1998), compact HII regions (C\'esarsky \etal 1996a) and AGNs
(Lutz \etal 1998; Roche \etal 1985).  Such sources are thus obviously not major
contributors to the integrated spectra of normal galaxies.  

ISO-SWS spectra
with greater spectral resolution show AFEs with the same shape, 
a clear indication that they are spectrally resolved in our PHT-S
data.  The non-detection of the 3.3\um\ feature in individual spectra is not 
surprising, since it is known to
amount to 1\% or less of the luminosity carried by the mid-infrared
AFEs in ``Type A sources'' (Tokunaga 1997; Willner \etal 1977 for 
M82)  and would therefore be below the 1$\sigma$ level in our
individual spectra.

Finally, an important consequence of the invariant shape of the spectrum up
to 11\um\, is that the 10\um\, trough is best  interpreted as a gap between
AFE rather than a silicate absorption feature.  An absorption feature would
become more pronounced in galaxies with larger infrared-to-blue ratios, and
that is not observed (see also Sturm \etal 2000).

\section{The Average Spectrum} \label{sec3}

The seven objects in Table 1 are among 45 galaxies observed by ISO for the
Key Project,  
most of which display similar spectra, regardless of the relative sizes of 
aperture and galaxy.  The only significant exception are NGC~4418, which is
known to harbor an AGN, and NGC~5866, an early-type galaxy discussed in
detail by Lu \etal (2000). 
Thus the mid-infrared spectral shape varies little or only weakly with galaxy
attributes.  Relative to each other, various feature luminosities 
are constant to within the signal-to-noise ratio, or 
$\sim20\%$.  One exception is the relative strength of the 11.3\um\ feature
which varies among galaxies by as much as 40\%, to be  discussed in more
detail elsewhere (Lu \etal 1996, Lu \etal 2000). Figure~2 
shows a composite spectrum obtained by averaging the data from 28 galaxies,
including the seven in Figure~1, after
normalizing each spectrum to the integrated flux between 6 and 6.6\um.  
The 28 galaxies are a random subset of
the Key Project sample with diverse properties, ranging for instance
from 0.28 to 0.88 in $R(60/100)$.
This composite spectrum should be a reliable representation
of the emission from the ISM of  normal galaxies.  

The spectrum in Figure~2 is consistent with earlier data in this spectral
range, including early M82 spectra by Willner \etal (1977), ground-based
surveys (Roche \etal 1991), and IRAS-LRS data (Cohen \& Volk 1989).
However, it reveals new details and establishes the universality of the
AFE.  A striking aspect of the composite
spectrum is the smooth continuum stretching from 3 to 5\um, and apparently
underlying the AFE at longer wavelengths (see \S 5 below).

Madden (1996) and Boselli \& Lequeux (1996) show spectra of elliptical
galaxies dominated by stellar photosphere emission, which drop off between
2 and 5\um\, like $f_{\nu}\propto\lambda^{-2.5}\propto\nu^{2.5}$.  
This component appears negligible in the composite spectrum at
$\lambda\ge 3$\um, since the continuum has
$f_{\nu}\propto\nu^{0.65}$ at 3\um$\le\lambda\le$5\um.  
The well known 3.3\um \
aromatic feature is detected at the expected wavelength, carrying about
0.5\% of the power in the AFE longwards of 5\um, a significantly smaller
fraction than observed in M82 (Willner \etal 1977).

The small bump at 7\um \ is a
significant signal, carrying about 0.65\% of the total AFE power, and
might include the [Ar~II] $\lambda 6.985\um$ and the S(5)
pure rotational line ($v=0-0$) of H$_2$ $\lambda 6.910\um$.  In well
resolved ISO-SWS spectra, e.g. M82 or the line of sight to the Galactic
Center (Lutz \etal 1996), [Ar~II] clearly dominates.  The 0.65\% fraction
of AFE power, or about 0.1\% of the FIR luminosity, 
approaches the most luminous lines in the far infrared, [OI] and
[CII] (Malhotra \etal 1998).  Although no dust related feature has been
identified reliably at this wavelength, the high luminosity suggests that
such a feature may contribute in addition to the [Ar~II]+H$_2$ blend.
The smaller  bump at 9.6\um \ 
coincides with the  S(3) pure rotational line ($v=0-0$) of H$_2$ $\lambda
9.665\um$, but it is too luminous to be due to that line alone, 
scaling from well studied galaxies (Valentijn \etal 1996).  



\section{The Energetics} \label{sec4}

The fraction of starlight processed through AFE has been under debate since
the IRAS mission (Helou, Ryter \& Soifer 1991), and can now be directly
estimated using the new ISO data for the sample described above.  
The various AFE are measured by integrating the spectrum in the intervals 
6 to 6.5\um, 7 to 9\um, and 11 to 11.5\um.  The contribution from an
extrapolation of the 4\um\, continuum is completely negligible, below the 1\%
level.  The 11.3\um\
feature merges into a complex that extends to about 13\um.  We estimate the
total power of this complex by extending our composite spectrum 
using the mean, continuum-subtracted spectrum from Boulanger \etal (1996)
and C\'esarsky \etal (1996b).  This extension amounts to a 12\% adjustment
to the AFE emission within the PHT-S range.  
The result is that AFE
account for about 65\% of the total power between 3 and 13\um, and
about 90\% of the power between 6 and 13\um.
The AFE carry  25 to 30\% of L(FIR) in 
quiescent galaxies in our sample.  
This fraction gradually drops to less than 10\% 
in the most actively star forming
galaxies, i.e. those with the greatest L(IR)/L(B) ratio or 
$R(60/100)$, following the trend already noted in Helou, Ryter \& Soifer
(1991).  In a typical quiescent galaxy, AFE might carry 12\% of the total 
infrared dust
luminosity between 3\um \ and 1~mm, whereas all dust emission at
$\lambda<13\um$ come up to $\sim18$\% of the total dust luminosity.

In the individual galaxy spectra, the power integrated between 7 and 9\um \ 
runs at 2.5 to 3 times that between 6 and 7\um;
these integrals include the plateau between the features.  
From the composite spectrum, we find the integrals from 5.8 to 6.6\um, 
7.2 to 8.2\um, and 8.2 to 9.3\um \ to be in the ratio 1:2:1.

\section{The 4-Micron Continuum} \label{sec5}

Even though it lies an order of magnitude below the AFE peaks, the
continuum level shortward of 5\um \ is unexpectedly strong 
(see for instance the model of D\'esert \etal 1990).
The reliability of this continuum is not in question, since it was detected
in several individual galaxies with the same relative strength, and
was confirmed by ISO PHT-S staring observations of a few galaxies.  
However, the calibration of such weak signals may
be uncertain by more than the nominal 30\%; comparison with ISO-CAM data 
show a 30\% difference, with PHT-S on the high side.

The 4\um\, continuum flux density is positively correlated with the AFE
flux, strong evidence linking the continuum to dust rather than stellar
photospheres.  It appears to follow a power law $f_{\nu}\propto\nu^{+0.65}$
between 3 and 5\um, with an uncertainty of 0.15 on the power-law index.
Its extrapolation runs three times weaker than the observed $f_{\nu}(10\um)$,
leaving open the nature of the connection between the
4\um\, continuum and the carriers of the AFE.  Bernard \etal (1994) have
reported evidence for continuum emission from the Milky Way ISM in
COBE-DIRBE broad-band data at these wavelengths, with comparable amplitude.

Extrapolating the 4-\um \ continuum to longer wavelengths, and assuming 
the AFE are superposed on it, one finds that the
continuum contributes about a third of  the luminosity between 3 and 13\um, the
balance being due to AFE.  In the range between 6 and 13\um, that
fraction drops to about 10\%.  Against this extrapolated continuum, the
AFE, defined again as the emission from 5.8 to 6.6\um, 7.2 to 8.2\um, and
8.2 to 9.3\um , would have equivalent widths of about 4\um\ or $3.4\times
10^{13}$~Hz, 18\um\ or $9.2\times 10^{13}$~Hz, and 13\um\ or $4.9\times
10^{13}$~Hz, respectively.

The natural explanation for the 4-\um \ continuum is a population of 
small grains transiently 
heated by single photons to apparent temperatures $\sim1000$K.  Such a
population was invoked by Sellgren \etal (1984) to explain the 3\um\
emission in reflection nebulae, and by other authors to explain the IRAS
12\um\ emission in the diffuse medium (e.g. Boulanger \etal 1988).
Small particles with ten to a
hundred atoms have sufficiently small heat capacities that a single UV
photon can easily propel them  to 1000~K equivalent temperature
(Draine \& Anderson 1985).  Such a population is a natural extension of the
AFE carriers, though it is not clear from these data whether it is truly
distinct, or whether the smooth continuum is simply the non-resonant
emission from the AFE carriers.  While the current data cannot rule out
other contributions,
the shape does rule out a simple 
extension of the photospheric emission from main sequence stars.  Red
supergiants and Asymptotic Giant Branch stars may contribute,
though the level would have to be fortuitously
comparable to the dust emission at 10\um, and the superposition of emission
spectra would have to mimick a $\rm f_\nu\propto \nu^{+0.65}$ spectrum.

\section{Summary and Discussion}\label{sec6}

The mid-infrared spectra of normal star forming galaxies are dominated by
interstellar dust emission.  They are well described between 3 and 13\um \
as a combination of Aromatic Features in Emission and an underlying $\rm
f_\nu\propto \nu^{+0.65}$ continuum, with the Features carrying about 65\%
of  the
luminosity between 3 and 13\um.  One can reliably assume this is a
universal spectral signature of dust.

The constant spectral shape against changing heating conditions from
galaxy to galaxy is strong evidence for particles transiently
excited by individual photons rather than in  thermal equilibrium.
This explanation is
especially compelling because it accounts for both the aromatic features
and the continuum.  Transient heating obtains only within a
finite range of ISM phases, namely from the translucent molecular
regions, through the atomic, and up to weakly ionized regions.  In denser
regions, AFE carriers may be insufficiently illuminated 
(Beichman \etal 1988, Boulanger \etal 1990), or condensed onto
larger grains (Draine 1985), whereas in HII regions they would be destroyed
by ionizing radiation (C\'esarsky \etal 1996b).  Therefore
the AFE flux may be approximated as the integral over the appropriate 
ISM phases in each galaxy of the product of the AFE carrier 
cross-section and the heating intensity.
Helou, Ryter \& Soifer (1991) showed that the mid-infrared carries
a diminishing fraction of the dust luminosity as the star formation
activity increases in a galaxy.  While this has been interpreted as
the result of generally depressed AFE carrier abundance throughout the
whole galaxy, it is more likely to result primarily from relatively smaller
AFE carrier niches, presumably overtaken by HII
regions where harder and more intense radiation destroys AFE carriers.

The mid-infrared spectral shape is sufficiently uniform among galaxies that
it can be used for redshift determinations, using for instance the SIRTF
InfraRed Spectrometer (Roellig \etal 1998): $L(FIR)=10^9 \Lsun$ galaxies
can be readily detected out to z=0.1 in about an hour of integration time,
whereas 
ultraluminous galaxies may be detectable out to z=3 in a comparable amount
of time depending on the AFE-to-far-infrared ratio assumed (Weedman,
private communication).
For galaxies with known redshifts, AFE detection would be an unmistakable
dust signature, and thus instantly distinguish between thermal and
non-thermal mid-infrared emission, or quantify their relative importance
(Lutz \etal 1998).

\acknowledgments

We would like to thank L. Vigroux, X. D\'esert, F. Boulanger and 
B.T. Soifer for interesting discussions.
The anonymous referee's comments were helpful in improving the paper. 
GH acknowledges the
hospitality of IAS, Universit\'e Paris Sud, and of Delta Airlines during
part of this work.  The Infrared Space Observatory (ISO) is an ESA project
with instruments funded by ESA Member States (especially the PI countries:
France, Germany, the Netherlands and the United Kingdom), and with
participation by NASA and ISAS.  The ISOPHOT data presented in this paper
were reduced using PIA, which is a joint development by the ESA
Astrophysics Division and the ISOPHOT consortium.  This work was supported
in part by the Jet Propulsion Laboratory, California Institute of
Technology, under a contract with the National Aeronautics and Space
Administration.

%
%

\newpage

\begin{figure}
\plotone{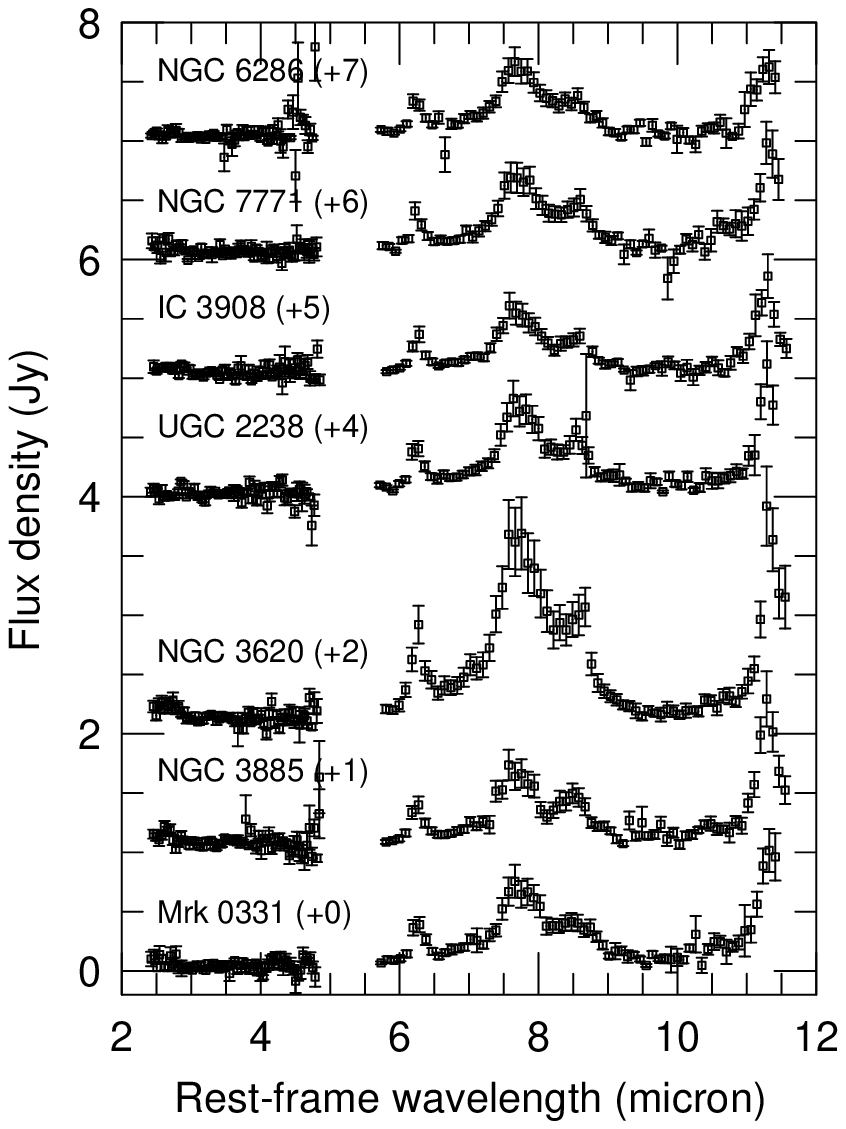}
\vspace{-0.7in}
\caption{ISO Mid-infrared spectra of the 7 galaxies listed in Table~1.
	 The ordinate is the  flux density within the PHT-S aperture of $24''\times
	 24''$, and the abscissa is the rest-frame wavelength in microns.
	 Each spectrum has been vertically shifted by a constant
	 given in the parentheses next to the name of the galaxy.
         }
\end{figure}
\newpage

\begin{figure}
\plotone{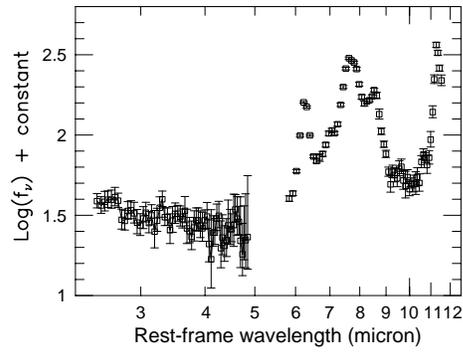}
\vspace{-2.5in}
\caption{Average spectrum obtained from a set of 28 galaxies, including the galaxies in
         Figure~1 (\S 3).   The quantities plotted
	 are the same  as those in Fig.~1, except for the use of
       logarithmic scales on both axes.  Error bars indicate the dispersion
among the averaged spectra when they are all normalized as described in the
text.
         }
\end{figure}
\newpage

\begin{deluxetable}{lclcccl}
\scriptsize
\tablenum{1}
\tablewidth{0pt}
\tablecaption{Galaxies with Mid-IR Spectra Shown in Figure 1}
\tablehead{
\colhead{Galaxy}  & \colhead{$R(60/100)$} & \colhead{Morphology} & 
\colhead{f$_{aper}$} & \colhead{L(FIR)/L(B)}  & \colhead{$\log\,$[L(FIR)/L$_{\sun}$]} &
\colhead{Mid-IR Morphology} \\
\colhead{(1)}  & \colhead{(2)} & \colhead{(3)} & 
\colhead{(4)}  & \colhead{(5)} & \colhead{(6)} & \colhead{(7)}  }
\startdata
NGC~6286       & 0.32  &  \ \ Sb   & 0.69       &    9   &  10.3 & inclined disk \nl
NGC~7771       & 0.47  &  \ \ SBa  & 0.52	&    7   &  10.4 & warped disk \nl
IC~3908        & 0.52  &  \ \ SBd  & 0.53	&    5   &   9.0 & inclined disk \nl
UGC~02238      & 0.52  &  \ \ Im?  & 0.80	&   11   &  10.4 & just resolved $\sim15^{\prime\prime}$ \nl
NGC~3620       & 0.68  &  \ \ SBab & 0.75	&  100   &  10.0 & faint disk, central peak \nl
NGC~3885       & 0.75  &  \ \ S0/a & 0.82	&    3   &   9.4 & just resolved $\sim15^{\prime\prime}$ \nl
Mkn~0331       & 0.81  &  \ \ S?   & 0.76	&   27   &  10.5 & just resolved $\sim15^{\prime\prime}$ \nl
\enddata
\end{deluxetable}

\begin{references}
\reference{} Allamandola, L. J, Tielens, A. G. G. M., \& Barker, J. R. 1989,
             \apjs, 71, 733
\reference{} Beichman, C.A., Wilson, R.W., Langer, W.D., \& Goldsmith,
             P.F. 1988, ApJL, 332, 81.
\reference{} Bernard, J.-P., Boulanger, F., D\'esert, F.X., Giard, M.,
             Helou, G. \& Puget, J.-L. 1994, \aap, 291, L5
\reference{} Boulanger, F., Falgarone, E., Puget, J.-L., \& Helou, G. 1990, 
ApJ, 364, 136.
\reference{} Boulanger, F. \etal 1988, ApJL, 332, 328.
\reference{} Boulanger, F. \etal 1996, A\&AL, 315, L325.
\reference{} C\'esarsky, C. \etal 1996, A\&AL, 315, L32.
\reference{} C\'esarsky, D. \etal 1996a, A\&AL, 315, L305.
\reference{} C\'esarsky, D. \etal 1996b, A\&AL, 315, L309.
\reference{} Cohen, M., \& Volk, K. 1989, \aj, 98, 1563.
\reference{} D\'esert, F.X., Boulanger, F. \& Puget, J.L. 1990, \aap, 273, 315.
\reference{} Draine, B.T. 1985, {\it Protostars and Planets II},
ed. D.C. Black \& M.S. Matthews (Tucson: University of Arizona), p. 621.
\reference{} Draine, B.T, \& Anderson, N. 1985, \apj, 292, 494.
\reference{} Geballe, T.R. 1997, {\it From Stardust to Planetesimals:
             Contributed Papers}, ed.  M.E. Kress, A.G.G.M. Tielens, \&
             Y.J. Pendleton, NASA Conference publication 3343 (Moffett
             Field: NASA ARC), p. 119. 
\reference{} Helou, G., Khan, I., Malek, L. \& Boehmer, L. 1988, ApJS, 68, 151.
\reference{} Helou, G., Ryter, C. \& Soifer, B.T. 1991, ApJ, 376, 505.
\reference{} Helou \etal 1996, A\&AL, 315, L157.
\reference{} Kessler, M.F. \etal 1996, A\&AL, 315, L27.
\reference{} Lemke, D. \etal 1996, A\&AL, 315, L64.
\reference{} Lu, N., Helou, G., Beichman, C.A. \etal 1996, BAAS, 28, 1356.  
\reference{} Lu, N. Y., \etal, 2000 (in preparation).
\reference{} Lutz, D. \etal 1996, \aap, 315, L269.
\reference{} Lutz, D. Spoon, H.W.W., Rigopoulou, D., Moorwood, A.F.M. \&
             Genzel, R. 1998, \apj, 505, L103.
\reference{} Malhotra, S., Helou, G., Stacey, G. \etal 1998, \apjl, 491, L27. 
\reference{} Metcalfe \etal 1996, A\&AL, 315, L105.
\reference{} Puget, J.-L. \& L\'eger, A. 1989, \araa, 27, 161.
\reference{} Roche, P.F., Aitken, D.K., Phillips, M.M. \& Whitmore, 
	     B. 1985, \mnras, 207, 35.
\reference{} Roche, P.F., Aitken, D.K., Smith, C.H. \& Ward, .J. 1991,
	     \mnras, 248, 606.
\reference{} Roellig, T.L., Houck, J.R., Van Cleve, J. \etal 1997, SPIE.
\reference{} Sellgren, K. 1984, \apj, 277, 623.
\reference{} Silbermann, N. \etal 1996, BAAS, 189, 6701.
\reference{} Silbermann, N. \etal 2000, in preparation.
\reference{} Sturm, E. \etal, 2000, in preparation
\reference{} Tokunaga, A.T. 1997, in {\it From Stardust to Planetesimals:
             Contributed Papers}, ed.  M.E. Kress, A.G.G.M. Tielens, \&
             Y.J. Pendleton, NASA Conference publication 3343 (Moffett
             Field: NASA ARC).
\reference{} Tuffs, R. 1998, in preparation.
\reference{} Valentijn, E.A., van der Werf, P.P., de Graauw, T. \& de Jong,
             T. 1996, \aap, 315, L145.
\reference{} Vigroux, L., \etal 1996, \aap, 315, L93.
\reference{} Willner, S., Soifer, B., Russell, R., Joyce, R., 
	     \& Gillett, F. 1977, \apj, 217, L121.
\end{references}
\end{document}